\documentclass[12pt,preprint]{aastex631}
\usepackage{multirow,amsthm,amsmath,amssymb,mathrsfs,ulem,url}
\usepackage{graphicx,subfigure,verbatim,color,soul}

\def\aap{Astron.\ Astrophys.\ }

\def\apjl{Astrophys.\ J.\ Lett.\ }
\def\apjs{Astrophys.\ J.\ Supp.\ }

\graphicspath{{fig/}}

\begin{document}

\title{Detection of afterglow emission up to 100 GeV through a stacking analysis of gamma-ray bursts}

\author{Shi Chen}
\affiliation{School of Physics and Astronomy, Yunnan University, Kunming 650091, P. R. China}
\affiliation{State Key Laboratory of Particle Astrophysics, Institute of High Energy Physics, Chinese Academy of Sciences, Beijing 100049, P. R. China}

\author{Qiang Yuan}
\affiliation{Division of Dark Matter and Space Astronomy, Purple Mountain Observatory, Chinese Academy of Sciences, Nanjing 210023, P. R. China; yuanq@pmo.ac.cn}
\affiliation{School of Astronomy and Space Science, University of Science and Technology of China, Hefei 230026, P. R. China}

\author{Yi-Qing Guo}
\affiliation{State Key Laboratory of Particle Astrophysics, Institute of High Energy Physics, Chinese Academy of Sciences, Beijing 100049, P. R. China}
\affiliation{University of Chinese Academy of Sciences, Beijing 100049, P. R. China}
\affiliation{TIANFU Cosmic Ray Research Center, Chengdu 610000, P. R. China; guoyq@ihep.ac.cn}

\author{Ben-Zhong Dai}
\affiliation{School of Physics and Astronomy, Yunnan University, Kunming 650091, P. R. China}

\author{He Gao}
\affiliation{School of Physics and Astronomy, Beijing Normal University, Beijing 100875, P. R. China}
\affiliation{Institute for Frontiers in Astronomy and Astrophysics, Beijing Normal University, Beijing 102206, P. R. China}

\author{Bing Zhang}
\affiliation{The Hong Kong Institute for Astronomy and Astrophysics, University of Hong Kong, Pokfulam Road, Hong Kong 999077, P. R. China}
\affiliation{Department of Physics, University of Hong Kong, Pokfulam Road, Hong Kong 999077, P. R. China}


\begin{abstract}
High-energy gamma-ray ($>$GeV) emission of gamma-ray bursts (GRBs) is very 
important in probing the jet evolution and particle acceleration of GRBs. 
The observations of high-energy photons are limited except for a few very 
bright GRBs, hindering precise measurements of the spectral and temporal 
evolutions of GRBs. Here we report the detection of high-energy gamma-ray 
emission up to 100 GeV with Fermi-LAT using a stacking analysis of a 
collection of 330 GRBs. High significance detection of the emission has 
been found, and the precise light curves and energy spectra can be measured. 
The light curves and time-resolved spectra of the sub-sample of 220 LAT 
individually detected GRBs can be well explained by the standard afterglow 
emission from a population of GRBs with both synchrotron and synchrotron 
self-Compton mechanisms, assuming a distribution of initial Lorentz factors. 
However, the emission of the relatively weak sample of the 110 LAT individually
undetected GRBs cannot be well reproduced in the same framework, indicating 
the existence of possible energy injection effect in the GeV band for the 
first time. The observations hence provide new insights in understanding 
the high-energy emission of GRBs.
\end{abstract}

\section*{}

GRBs are among the most extreme cosmic explosions, serving as unique 
laboratories for studying relativistic jets and particle acceleration 
under extreme conditions \citep{Zhang:2003uk,Kumar:2014upa}. While the prompt 
emission of GRBs is expected to be produced by internal shocks within the jet, 
the long-lasting afterglow arises from external shocks due to the jet 
interaction with the circum-burst medium which then produces wide-band 
emission via synchrotron and the inverse Compton scattering processes
\citep{Sari:2000zp,Zhang:2010ey}. The high-energy emission of GRBs is 
usually dominated by this afterglow component \citep{Ajello:2019zki}, 
and thus plays a crucial role in constraining the jet parameters, magnetic 
field properties, and the maximum accelerated energies for GRBs
\citep{Kumar:2014upa,MAGIC:2019irs}. However, observations of high-energy 
emission are rare, due to the limited effective area of space detectors and 
the challenges of groundbased observations (small field-of-view or high 
background). Currently, only a few very bright GRBs have been detected to 
have enough photons enabling a good description of the spectra and light 
curves by the space detector Fermi Large Area Telescope (LAT)
\citep{Ajello:2019zki,Fermi-LAT:2022byn} and groundbased experiments
\citep{MAGIC:2019lau,HESS:2021dbz,LHAASO:2023kyg}. This situation cannot 
be easily improved in the near future for individual GRB observations. 
A stacking analysis for a population of sources may effectively overcome 
this issue, as has been done for pulsars \citep{McCann:2014dea}, galaxy 
clusters \citep{Huber:2013cia}, and the prompt phase of GRBs
\citep{Fermi-LAT:2016gue}\footnote{Note that in \citet{Fermi-LAT:2016gue} 
79 GRBs with Swift localizations were investigated with a stacking technique. 
However, no detailed spectral and temporal behavirors of the GRB sample were 
studied.}. 

In this work we employ the stacking technique to study the high-energy 
gamma-ray emission from a sample of GRBs using the data recorded by the 
Fermi-LAT instrument \citep{Fermi-LAT:2009ihh}. From August, 2008 to December, 
2024, a total of 222 GRBs have been reported by Fermi-LAT, triggered by either 
the LAT or the Gamma-ray Burst Monitor (GBM) \citep{Meegan:2009qu}. In our 
analysis, 220 GRBs out of the 222 ones, excluding two bright GRBs 130427A
\citep{Fermi-LAT:2013ufa} and 221009A \citep{Fermi-LAT:2024jox}, are used. 
These LAT individually detected GRBs (``LAT detected 
GRBs'' for short) exhibit diverse properties of energy distributions and 
other statistical properties \citep{Ajello:2019zki}. For most of these GRBs 
the numbers of high-energy photons are very few, for which the spectra and 
light curves can not be obtained and thus the high-energy behaviors of the 
emission can not be well extracted. We also compile a sample of 110 GRBs 
which were triggered by Swift/BAT \citep{Donato:2012xx} and not detected 
by Fermi-LAT (referred to as ``LAT individually undetected GRBs'' and 
``LAT undetected GRBs'' for short), with a filtering condition that at 
least one event exists within $1^{\circ}$ radius of each GRB. Such a 
sample represents the faintest end of GRBs can be studied with the current 
Fermi-LAT detector.

\begin{figure}[thb]
\centering
\includegraphics[width=0.49\textwidth]{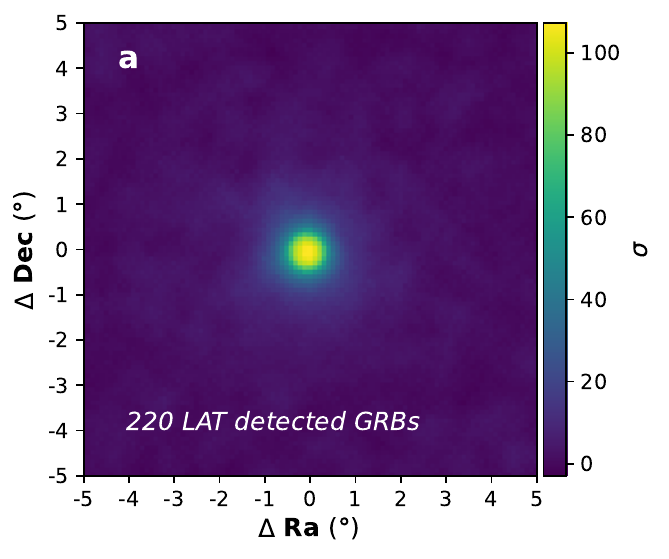}
\includegraphics[width=0.48\textwidth]{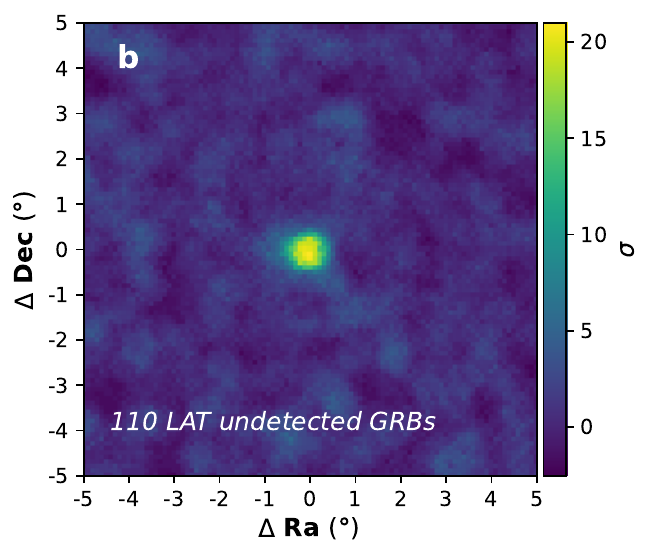}
\caption{The $10^{\circ} \times 10^{\circ}$ significance maps of the 
stacked GRBs smoothed with a $0.4^{\circ}$ Gaussian kernel. For each GRB, 
photons in the region of the same size centered on its trigger position 
within a time duration from the trigger time $T_0$ to $T_0 + 50,000$ s are 
collected. Panel {\bf a} is for the 220 LAT detected GRBs with photon energies 
from 0.1 to 100 GeV, and panel {\bf b} is for the 110 LAT undetected GRBs 
with energies from 1 to 10 GeV.}
\label{fig:sig}
\end{figure}

We use the official Fermi Science Tools provided by the Fermi Science 
Support Center (FSSC) to analyze the data. Photons in the $0.1-100$ GeV 
band from a $10^{\circ}\times10^{\circ}$ region centered on the location 
of each GRB for a time window of 50000 seconds after the trigger have 
been stacked. The background contains emission from the Galactic diffuse 
background, the isotropic diffuse background, and sources from the 4FGL 
catalog \citep{Fermi-LAT:2022byn} (see Appendix A for more details). 
Figure \ref{fig:sig} shows the significance maps of the 220 LAT detected 
sample (panel {\bf a}) and the 110 LAT undetected sample (panel {\bf b}). 
Note that, for the 110 LAT undetected sample, the number of photons is 
limited and only significant detection in the $1-10$ GeV is found and 
hence the significance map is derived using the $1-10$ GeV events. 
Significant detection of both samples can be found, with significance 
of $99.7\sigma$ for the 220 LAT detected sample and $17.9\sigma$ for 
the 110 LAT undetected sample (see Appendix B for the calculation of 
the significance). To check that whether these excess photons indeed 
come from GRBs, we derive the one-dimensional angular distribution of 
the photons after subtracting the background, for the 220 LAT detected 
sample. The results in different energy bands can be modelled with 
Gaussian functions, as shown in Figure \ref{fig:angular} of Appendix. 
The angular distributions are consistent with the point spread function 
(PSF), as indicated by the standard pointlike source of the crab (including 
both the pulsar and nebula), except in the lowest energy band where different 
spectral shapes may result in slightly different PSFs. 

\begin{figure}[thb]
\centering
\includegraphics[width=0.49\linewidth]{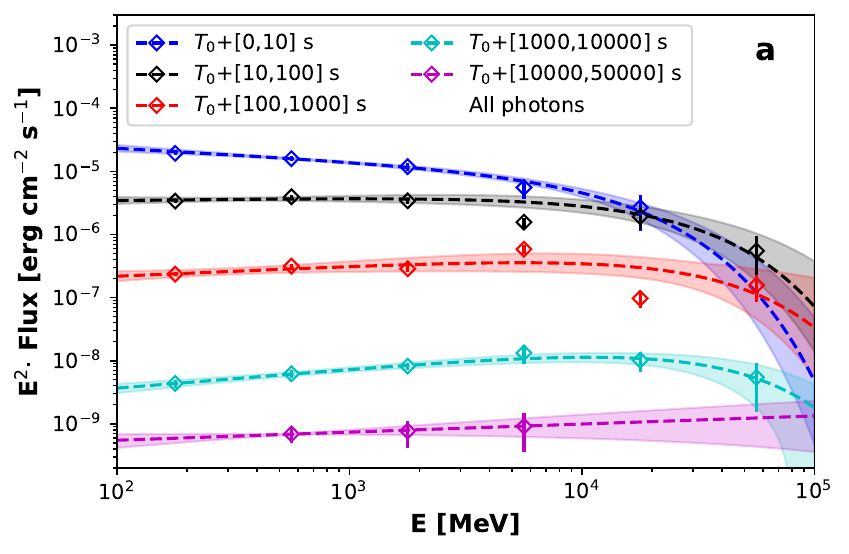}
\includegraphics[width=0.49\linewidth]{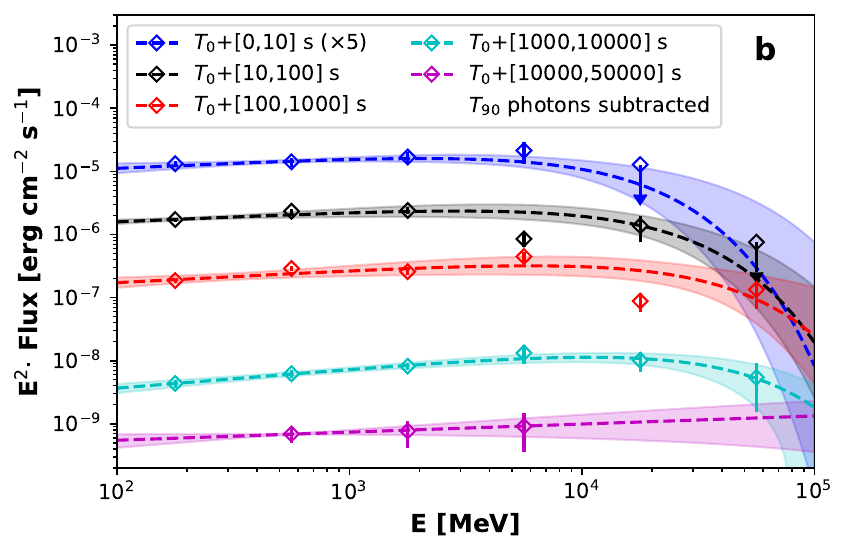}
\caption{SEDs at different time for the 220 LAT detected GRBs. Five time 
intervals are adopted: $T_0+[0,10]$ s, $T_0+[10,100]$ s, $T_0+[100,1000]$ s, 
$T_0+[1000,10000]$s, and $T_0+[10000,50000]$ s. Panel {\bf (a)} shows the 
results obtained for all photons, and panel {\bf (b)} shows the results 
after removing photons within $T_{90}$ of each GRB. Dashed lines and shaded 
bands show the best-fitting results and the $\pm1\sigma$ uncertainty bands 
with an ECPL model.}
\label{fig:spec}
\end{figure}

We extract photons in different time windows after their trigger time 
($T_0$), and derive the spectral energy distributions (SEDs) at different 
time. The results of the 220 LAT detected GRBs are shown in Figure 
\ref{fig:spec}. One can see that if all the photons are taken into account 
(panel {\bf a}), the spectra experience a clear evolution from soft at early 
time to hard at late time. However, there might be contribution from prompt 
emission at early time. To reduce the possible impacts from the prompt 
emission, we show in panel ({\bf b}) the SEDs after subtracting photons 
within $T_{90}$ of each GRB, where $T_{90}$ is defined as the time interval 
containing $90\%$ of the $\gamma$-ray photons during the burst phase. 
The SEDs appear similar among different time. We fit the SEDs with an 
exponential cutoff power-law (ECPL) model, 
$\phi(E)\propto E^{-\alpha}\exp(-E/E_{\rm cut})$. The evolution of spectral 
indices with time is given in Figure \ref{fig:index} (panel {\bf a}). 
We can see that if the photons within $T_{90}$ are removed, the spectral 
index keeps roughly a constant of $\alpha \sim 1.8$. Including the photons 
within $T_{90}$ results in softer spectra at early time
\citep{Ravasio:2023vgh}. The cutoff energy shows a trend of increase with 
time (panel {\bf b} of Figure \ref{fig:index}), which may indicate that the 
high-energy component from the inverse Compton scattering process becomes 
more and more important at later time (see below discussion on the modeling).

\begin{figure}[thb]
\centering
\includegraphics[width=0.49\textwidth]{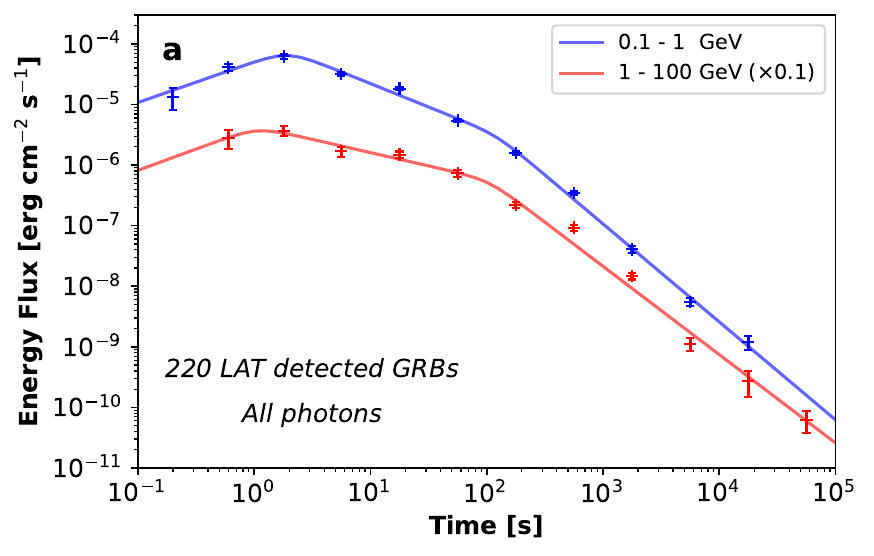}
\includegraphics[width=0.49\textwidth]{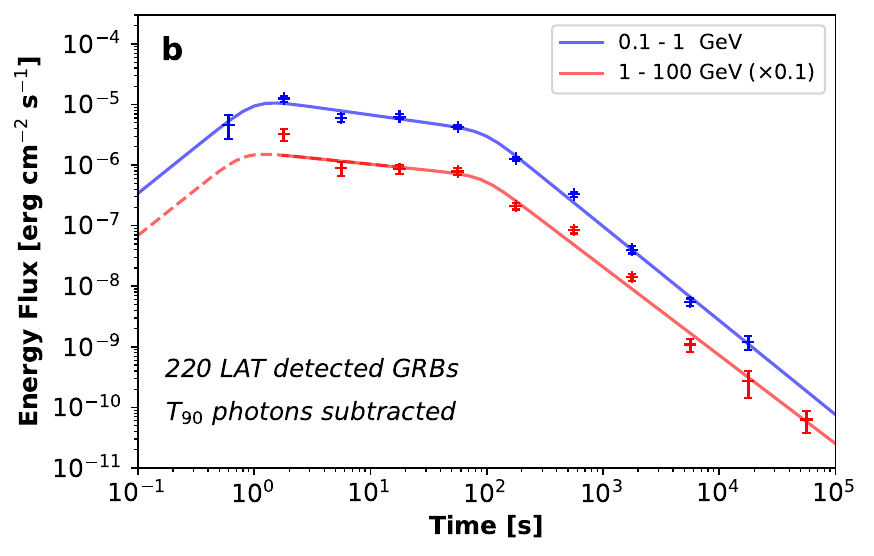}
\includegraphics[width=0.49\textwidth]{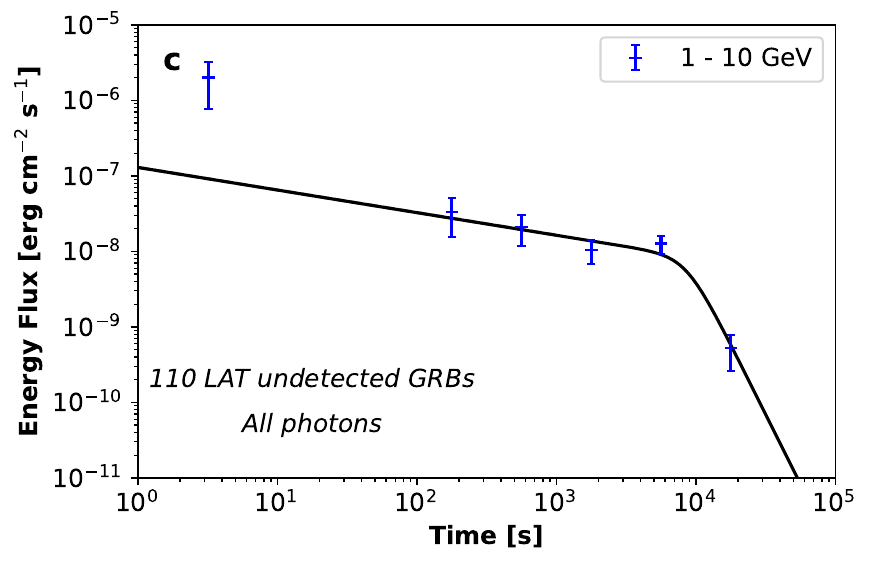}
\includegraphics[width=0.49\textwidth]{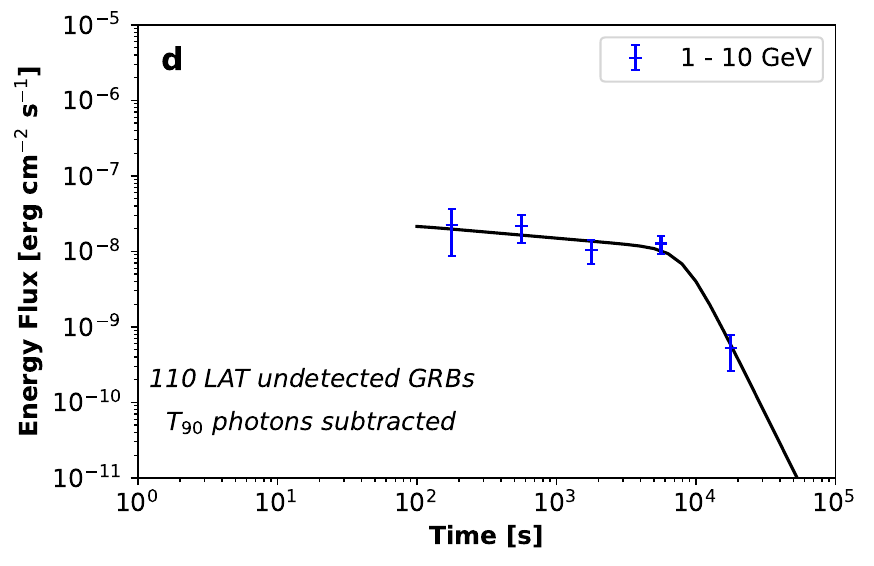}
\caption{Light curves of the stacked GRBs and the corresponding fitting 
results. The light curves for all photons and for those with $T_{90}$ 
photons subtracted are shown in panels {\bf a} and {\bf b} for the 220 
LAT detected GRBs, and in panels {\bf c} and {\bf d} for the 110 LAT 
undetected GRBs, respectively. All light curves are fitted using a 
multi-piece power-law model, with fitting parameters being given in 
Table \ref{tab:lc}.}
\label{fig:lc}
\end{figure}

The light curves of the two samples are shown in Figure \ref{fig:lc}, 
for all photons (panels {\bf a} and {\bf c}) and photons within $T_{90}$ 
subtracted (panels {\bf b} and {\bf d}). For the 220 LAT detected GRBs, 
the results in two energy bands, $0.1-1$ GeV and $1-100$ GeV, are presented 
(panels {\bf a} and {\bf b}), while for the 110 LAT undetected GRBs the 
results in $1-10$ GeV are shown (panels {\bf c} and {\bf d}). The light 
curves of the 220 LAT detected GRBs can be fitted with three-piece power-law 
functions, as shown by solid lines in the figure. For the case with all 
photons, the light curves show a rise at the early time with a slope of 
$\sim 0.7$, followed by a shallow decay (with decay slope of $-0.8\sim-0.4$) 
after a few seconds and a steep decay (with decay slope of $\sim -1.5$) at 
about 100 s after the trigger. The slopes of the shallow decay phase vary 
with energies, which may be due to the prompt emission contamination. 
If we remove the photons within $T_{90}$, the light curves of different 
energy bands become similar with each other. In this case, the broken 
power-law evolution behaviors are also visible, but the rising phase 
becomes less clear due to limited number of photons. The slopes of the 
shallow decay phase are about $-0.2\sim-0.3$, which evolve to the steep 
decay phase with slope of $\sim -1.5$ at $\sim100$ s after the trigger. 
For the 110 LAT undetected GRBs, a two-piece power-law behavior can be 
seen, but the rising phase as shown in the 220 LAT detected sample cannot 
be detected. The slope of the shallow decay is again about $-0.2\sim-0.3$, 
followed by a steep decay with slope of $-3.7$ but with relatively large 
uncertainty. The break time from the shallow decay to the steep decay is 
different for the two samples ($\sim100$ s for the 220 GRBs and $\sim10^4$ 
s for the 110 GRBs), which we expect may be due to different Lorentz factors 
of the jets and/or late-time energy injection effect for the two samples.

In the standard afterglow model of GRBs \citep{Sari:1997qe,Kumar:2009vx}, 
the slow rise phase may correspond to the coasting stage that the forward 
shock sweeps up the ambient medium before significant deceleration. After 
that, the jet enters the deceleration phase, and the light curve starts 
to decay. With the continuous deceleration of the jet, when the Lorentz 
factor of the jet becomes smaller than $1/\theta_j$ where $\theta_j$ is the 
half-opening angle of the jet (known as the jet break \citep{Rhoads:1999wm}), 
a steeper decay of the emission is expected. Eventually, the jet becomes 
non-relativistic and enters the Newtonian evolution stage. Depending on 
the specific parameters, the evolution slopes of the light curves differ 
case by case \citep{Dermer:1999eh,Gao:2013mia}. The results obtained in 
this work likely show the expected evolution of the standard afterglow 
model. However, a more complicated situation here is the cumulative effect 
of many GRBs with diverse properties.

\begin{figure}[!thb]
\centering
\includegraphics[width=0.48\textwidth]{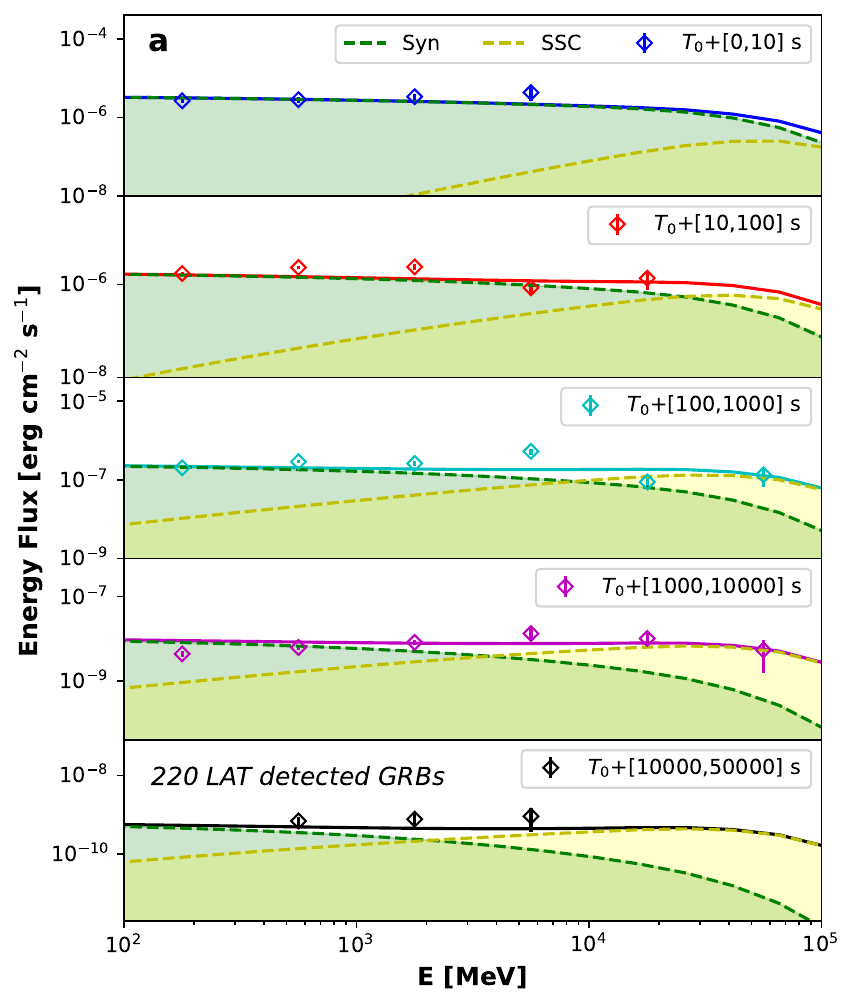}
\includegraphics[width=0.48\textwidth]{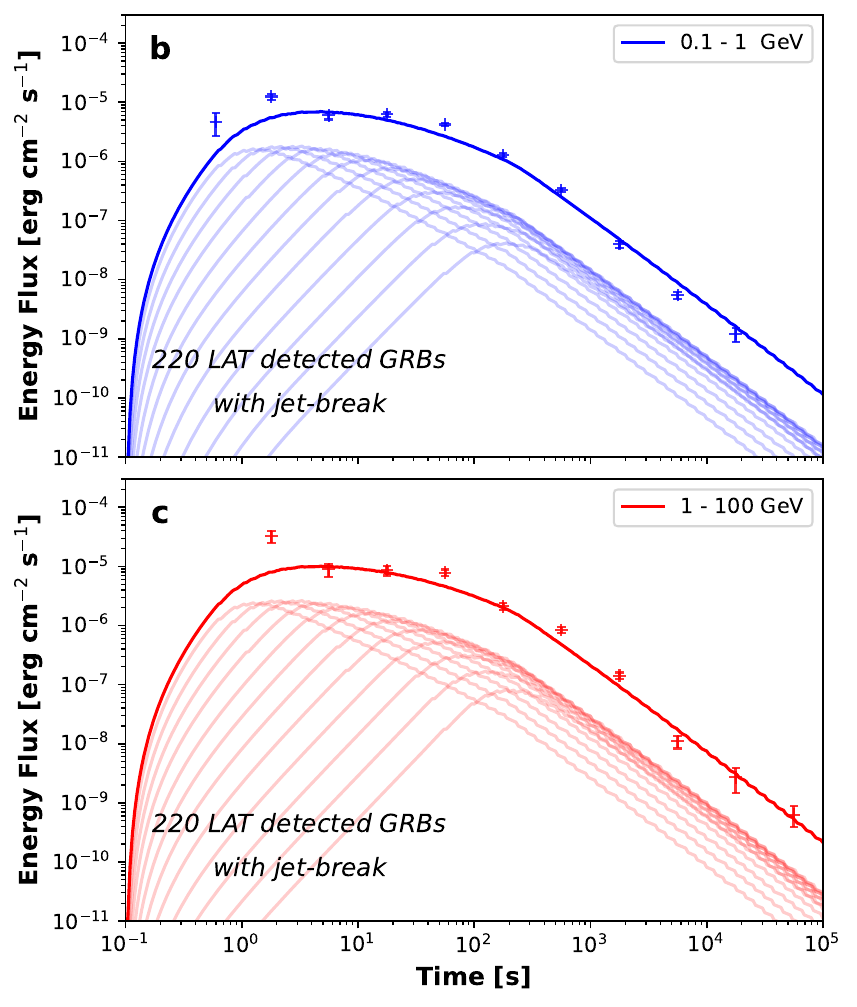}
\includegraphics[width=0.48\textwidth]{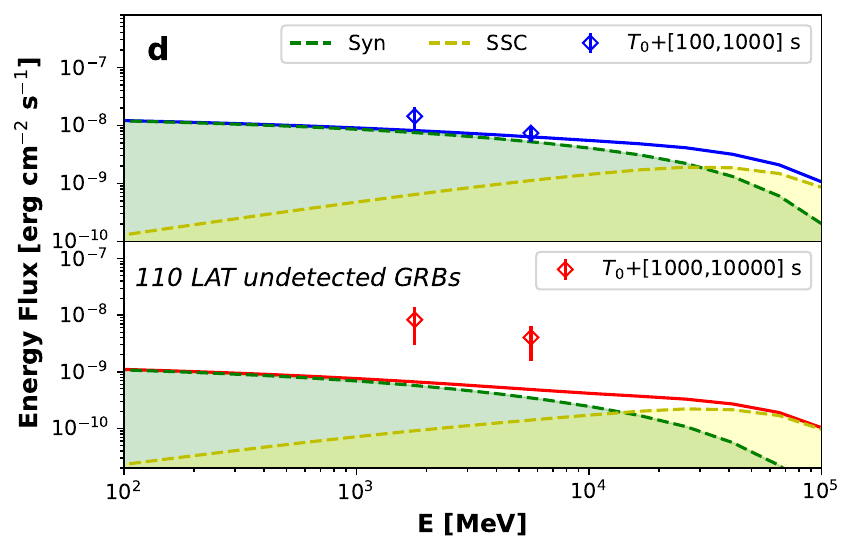}
\includegraphics[width=0.48\textwidth]{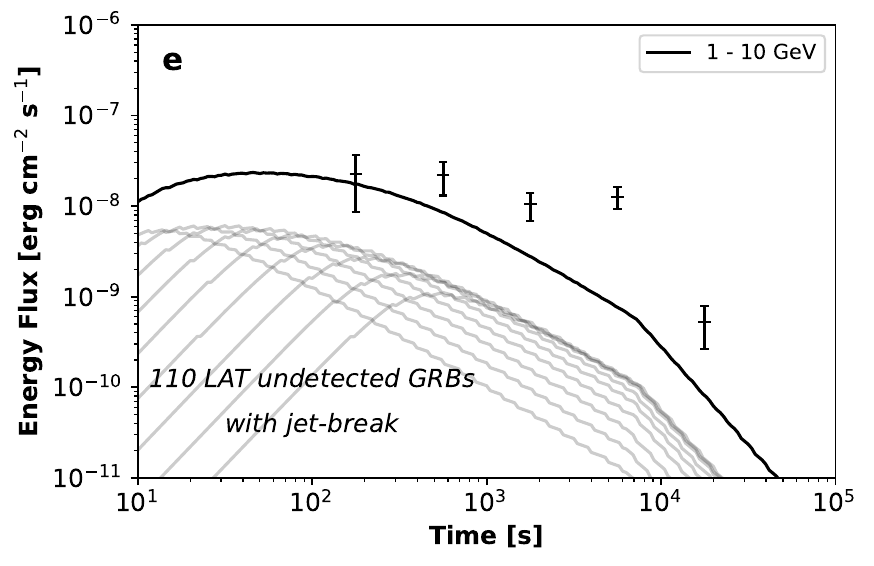}
\caption{Stacked SEDs and light curves from GRB population with log-normal 
initial Lorentz factor distribution. Panels {\bf a}, {\bf b} and {\bf c} 
present the stacked model SEDs and light curves for the high initial Lorentz 
factor ($\Gamma_0 \sim 130-800$) populations, corresponding to the 220 LAT 
detected GRBs. Panels {\bf d} and {\bf e} show the corresponding stacked 
model SEDs and light curves for the low initial Lorentz factor ($\Gamma_0 
\sim 60-250$) populations, which are derived for the 110 LAT undetected GRBs.
}
\label{fig:model}
\end{figure}

We model the SEDs and light curves with the forward shock model of the 
GRB afterglow. In this model, the propagation of relativistic jets in the 
circum-burst medium produces forward shocks which accelerate high-energy 
electrons. These energetic electrons then produces multi-wavelength emission 
through synchrotron and/or inverse Compton emission. In the energy band of 
Fermi-LAT, both the synchrotron emission and the synchrotron self-Compton 
(SSC) emission are relevant \citep{Zhang:2001az,Wang:2019zbs}. The parameters 
of the GRBs in our sample should be different from one to another. One of 
the major parameter affecting the afterglow evolution is expected to be 
the initial Lorentz factor of the jet, $\Gamma_0$. The initial kinetic 
energy ($E_0$) of the jet which affects the normalization of the emission 
is also a highly relevant parameter. Following the results of 
\citet{Liang:2009zi}, we assume that the initial Lorentz factor has a 
log-normal distribution, with the mean value of $\log(\Gamma_0)=2.37$ and 
the width of $0.26$. For the 220 LAT detected sample, which are expected 
to be more powerful, $E_0$ is taken to be $4\times10^{53}$ erg, $\Gamma_0$ 
is restricted in the range of $[130,800]$, and the jet opening angle is 
assumed to be $0.8^{\circ}$. For the 110 LAT undetected sample, $E_0$ is 
assumed to be $4\times10^{52}$ erg, $\Gamma_0$ is restricted in the range 
of $[60,250]$, and the jet opening angle is taken to be $4^{\circ}$. 
The initial Lorentz factor and the kinetic energy follows also the 
correlation of $\Gamma_0\sim E_0^{0.25}$ \citep{Liang:2009zi}, as shown 
in Figure \ref{fig:lognormal} of the Appendix. The other parameters are 
fixed to be the same for all GRBs (see Appendix E for details). The 
stacked model results of the SED and light curves are given in Figure 
\ref{fig:model}. 

The model successfully reproduces the three-segment power-law evolution 
of the light curves for the 220 LAT detected GRBs. The emission reaches 
a peak after the rising phase, when the jet starts to decelerate. The 
peak time is related with the initial Lorentz factor as 
$t_{\rm peak}\propto \Gamma_0^{-8/3}$. Therefore, the shallow decay phase 
we observe can be ascribed to the superposition of various shallow decay 
curves for the population of GRBs. The shallow decay slope of $-0.2\sim-0.3$ 
depends on the distribution of the initial Lorentz factor, and is flatter 
than the shallow decay phase of individual GRB. The light curve enters the 
steep decay phase after the jet break of the GRB with the latest 
$t_{\rm peak}$. The jet break is necessary to reproduce the steep decay 
phase of the light curve. As can be seen in Figure \ref{fig:lognormal}, 
the data cannot be well accounted for without the jet break. 

For the 110 LAT undetected GRBs, the model prediction shows some tension 
with the observations. It is interesting to note that the break time from 
the shallow decay to the steep decay is about $10^4$ s, just consistent 
with the durations of X-ray plateaus of canonical GRBs \citep{Zhang:2005fa}. 
For the GeV bright GRBs detected by Fermi-LAT, the X-ray light curves usually 
show single power-law decay without clear plateaus \citep{Yamazaki:2019gez}. 
This is consistent with the results obtained here, i.e., long-lasting emission 
exists in the sample with weaker GeV emission. As a consistency test, we 
compile the available X-ray afterglow data of the two samples studied in 
this work, and find that for the 110 LAT undetected sample the X-ray light 
curves indeed tend to favor the existence of energy injection more than the 
220 LAT detected sample (see Appendix F). Our results likely give the first 
detection of GeV counterparts of canonical GRBs with X-ray plateaus, and 
indicate that later energy injection from the central engine may take effect 
in shaping the evolution of $\gamma$-ray afterglow for the class of weak GRBs.

The SEDs can be roughly explained by the model for the 220 LAT detected sample,
but show excesses for the 110 LAT undetected sample. Both the synchrotron and 
SSC components contribute to the time-resolved spectra and light curves, 
suggesting that the transition of radiation mechanisms occurs in the $O(10)$ 
GeV $\gamma$-ray band. The energy flux of the SSC component is comparable to 
the synchrotron component, which is very useful in constraining the energy 
allocation of GRB jets \citep{Joshi:2019opd} and the contribution to the 
extragalactic $\gamma$-ray background from the GRB population 
\citep{Casanova:2006bp}. Time evolution of these two components can be seen: 
the synchrotron component dominates the emission at early time, and with the 
increase of time the SSC component gradually becomes more and more important, 
especially in the high energy band. This can also explain the increase trend 
of the cutoff energy as obtained in the phenomenological ECPL fitting of the 
SEDs (see Figure \ref{fig:index} of the Appendix).

The results of this analysis show that the high-energy afterglow emission 
of GRBs should generally exist up to 100 GeV. However, for single event 
the number of photons is usually very limited for space detectors and the 
detection of such emission is very difficult. With the stacking technique, 
we can reach the high significance detection of such emission and precisely 
characterize the spectral and temporal properties. The results verify the 
canonical afterglow emission due to interactions of GRB jets with the 
circum-burst medium \citep{Ghisellini:2009rw}. It is also possible that 
there is energy injection signature in the GeV $\gamma$-ray band for the 
relatively weaker sample. The jet-break like feature has been observed on 
the light curves, which can provide useful constraints on the jet dynamics. 
Although it is not clear to what energy the prompt emission can extend, 
the SEDs indicate that possible prompt emission up to several GeV exists 
in the data. It is currently challenging to decompose the prompt emission 
and the afterglow within the bursting phase ($T_{90}$). Future high-energy 
$\gamma$-ray observations with bigger effective area \citep{Pan:2024adp} 
may better address this issue.

\section*{Acknowledgements}
This work is supported by the National Natural Science Foundation of China 
(Nos. 12321003, 12220101003, 12263007, and 12233006), the Project for Young 
Scientists in Basic Research of the Chinese Academy of Sciences (No. YSBR-061),
and the High-level Talent Support Program of Yunnan Province.

\appendix

\setcounter{figure}{0}
\renewcommand\thefigure{A\arabic{figure}}
\setcounter{table}{0}
\renewcommand\thetable{A\arabic{table}}

\section{Stacking analysis of the Fermi-LAT data}

From August, 2008 to December, 2024, 222 GRBs have been detected by Fermi-LAT, 
with detailed information being available through the 
GRBweb\footnote{\url{https://user-web.icecube.wisc.edu/~grbweb_public/index.html}}. 
We use the Fermitools from the Fermi Science Support Center (FSSC) to analyze 
the data. To enable a good statistics, the {\tt SOURCE} class events with 
reconstruction version of {\tt P8R3$\_$SOURCE$\_$V3} have been adopted. 
For each GRB, we select events within a region of interest (ROI) of 
$10^{\circ}$ centered on the location of the burst. Both {\tt FRONT} and 
{\tt BACK} events are adopted. The time range is chosen to be 50000 s after 
the trigger time $T_0$, and the energy range is from 100 MeV to 100 GeV. 
The maximum zenith angle is set to be $90^{\circ}$ to filter out the 
background from the Earth’s limb. The event selection further includes 
filters {\tt (DATA\_QUAL\textgreater0)} \&\& {\tt (LAT\_CONFIG$==$1)} to 
ensure good quality of the data.

A count map is derived for each GRB. By stacking the count maps of all 220 GRBs
excluding the very bright two GRBs 130427A and 221009A, we obtain a total 
count map. This map includes signal events from GRBs, and background emission 
from the isotropic gamma-ray background (IGRB) \citep{Fermi-LAT:2014ryh}, 
the diffuse Galactic background \citep{Fermi-LAT:2012edv}, and sources 
\citep{Fermi-LAT:2022byn} in the ROI. 

In addition, a sample of Swift/BAT triggered GRBs but not detected by 
Fermi-LAT individually from the same period of August, 2008 to December, 
2024 has also been collected. We further require that there is at least 
one photon like event within $1^{\circ}$ radius of each GRB after the 
above event selections. It turns out that 110 GRBs remain in this sample.

\section{Likelihood fitting}

We apply the likelihood analysis on each GRB. Usually the binned likelihood 
method is adopted, unless for very short time bins (less than 200 s) where 
the number of events is very limited and the unbinned likelihood analysis 
is adopted. The model includes the signal from the GRB, the sources in the 
ROI from the 4FGL catalog \citep{Fermi-LAT:2022byn} ({\tt gll\_psc\_v32.fit}), 
and the diffuse backgrounds\footnote{http://fermi.gsfc.nasa.gov/ssc/data/access/lat/BackgroundModels.html}. 
For the diffuse Galactic emission we use the template {\tt gll\_iem\_v07.fits}, 
and for the isotropic background we use the spectral model 
{\tt iso\_P8R3\_SOURCE\_V3\_v1.txt}. The background emission (including both 
the diffuse backgrounds and field sources) for each GRB is obtained and 
subtracted from the data. Excess emission from the sample of GRBs can be 
obtained. The significance map for 220 GRBs observed by LAT and 110 GRBs 
not recorded by LAT is shown in Figure \ref{fig:sig}. 

\begin{figure}[!htb]
\centering
\includegraphics[width=0.48\textwidth]{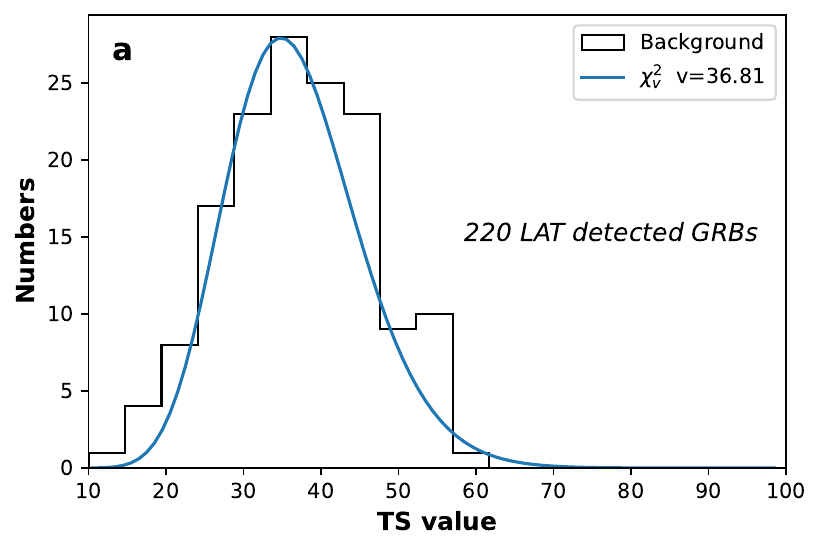}
\includegraphics[width=0.48\textwidth]{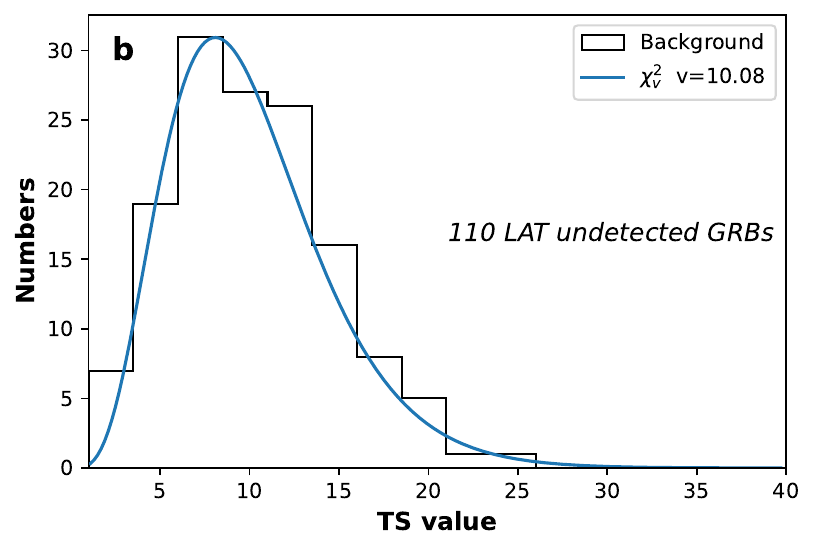}
\caption{The distributions of TS values of pseudo-experiments, for the 
220 LAT detected sample (panel {\bf a}) and the 110 LAT undetected sample 
(panel {\bf b}). The solid line in each panel is the best-fitting 
$\chi^2_{\nu}$ distribution with degree-of-freedom $\nu$.
}
\label{fig:TS}
\end{figure}

Summing the logarithmic likelihood values together, we get the total Test 
Statistic (TS) value, defined as ${\rm TS}=2(\ln{\mathcal L}-
\ln{\mathcal L}_0)$, where ${\mathcal L}$ and ${\mathcal L}_0$ represent 
the likelihood values with and without target GRBs, as 10181.8 for the 220 
LAT detected GRBs across $0.1-100$ GeV band from $T_0$ to $T_0+50,000$ 
seconds. For the 110 LAT undetected GRBs, the TS value is about 366.8 in 
the $1-10$ GeV band from $T_0$ to $T_0+50,000$ seconds. To obtain the 
detection significance, we carry out pseudo-experiments as follows. 
We adopt the same sky regions of the selected GRB samples but choose data 
three months after each burst, and bin the data into 50000 s bins, and re-do 
the same likelihood fitting to calculate the TS values of the pseudo-data. 
The distributions of the TS values are shown in Figure \ref{fig:TS}.
We find that the TS distributions can be well fitted using the $\chi^2_{\nu}$ 
distribution with degree-of-freedom $\nu$, and get $\nu=36.8$ and 10.1 for 
the two samples, respectively. According to the fitting $\chi^2_{\nu}$ 
distributions, we estimate that the detection significance of GRBs is 
approximately $99.7\sigma$ and $17.9\sigma$ for the LAT detected and 
undetected samples.

\begin{figure}[!htb]
\centering
\includegraphics[width=0.32\textwidth]{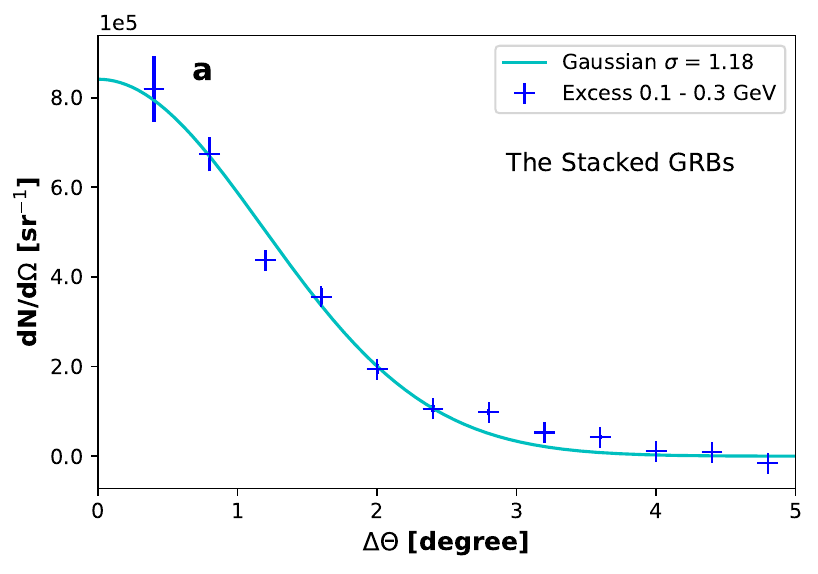}
\includegraphics[width=0.32\textwidth]{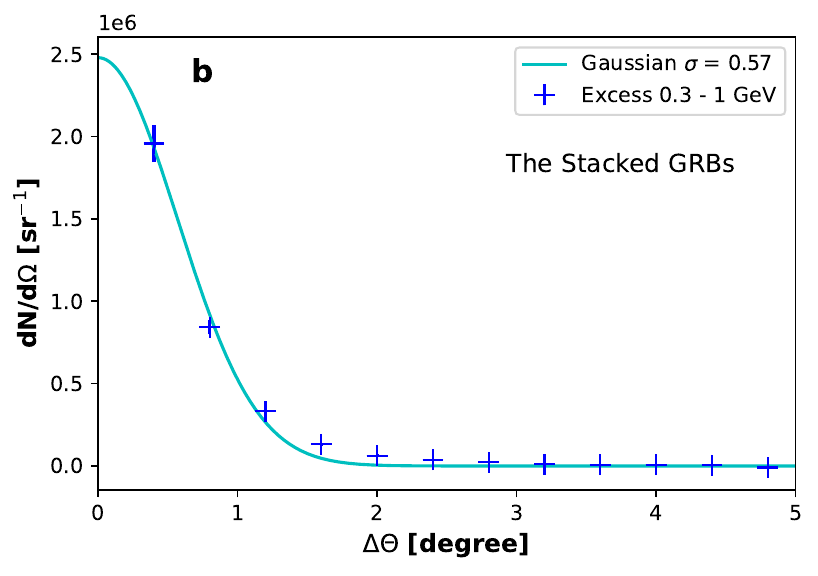}
\includegraphics[width=0.32\textwidth]{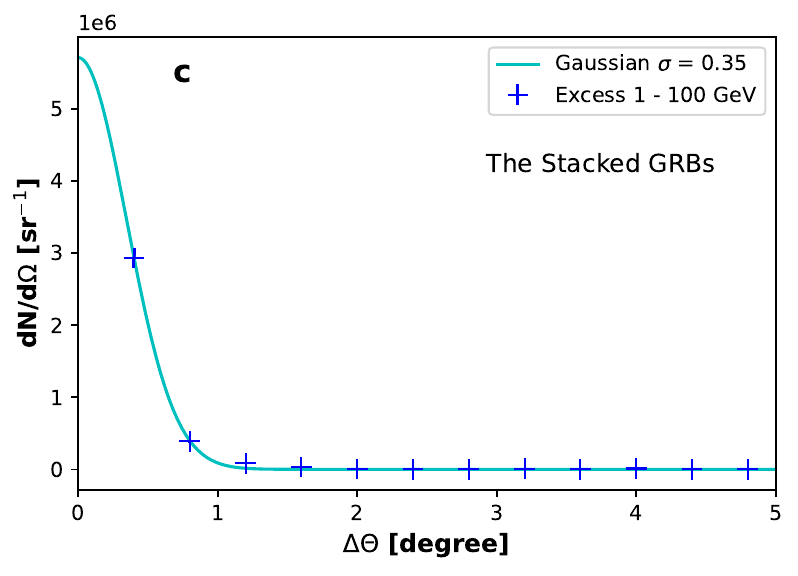}
\includegraphics[width=0.32\textwidth]{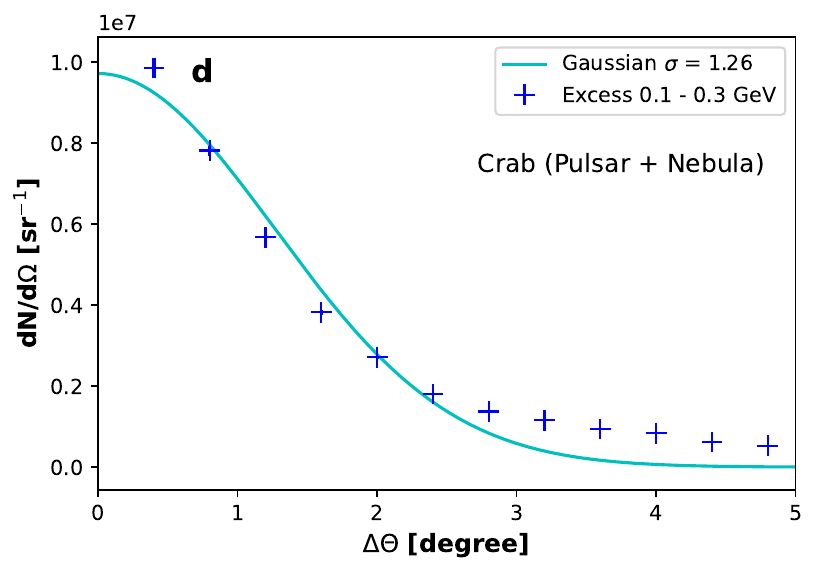}
\includegraphics[width=0.32\textwidth]{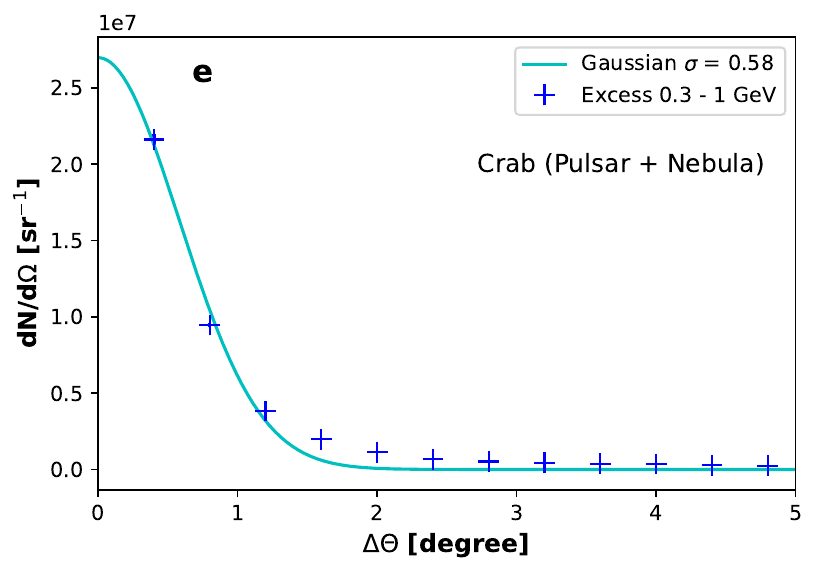}
\includegraphics[width=0.32\textwidth]{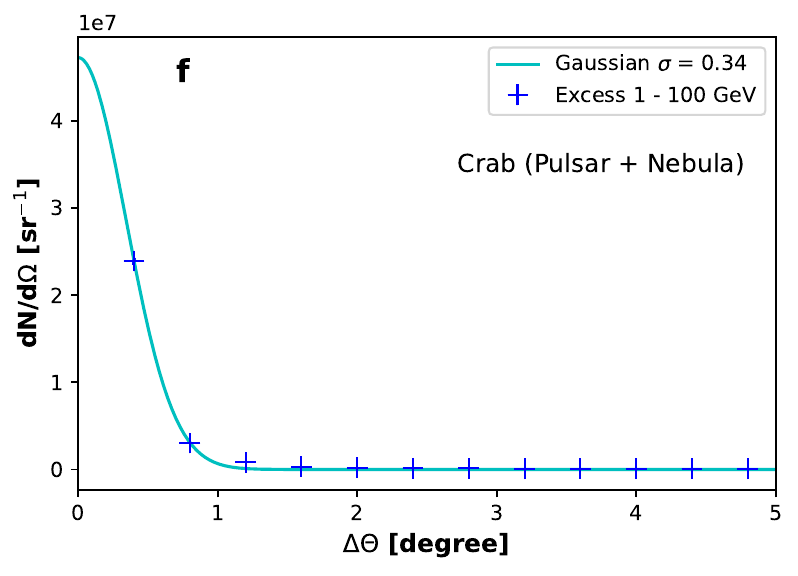}
\caption{Angular distributions of excess photons from stacked GRBs (panels 
{\bf (a)-(c)}) and from crab (panels {\bf (d)-(f)}). Results for three energy 
bands, $0.1-0.3$ GeV, $0.3-1$ GeV, and $1-100$ GeV, are shown. For the GRB 
sample, the data from $T_0$ to $T_0 + \text{50,000}$ s are extracted, and 
for crab the data of one year observation (from August 4, 2008 to August 4, 
2009) are used. The solid line in each panel shows the Gaussian fitting result.
}
\label{fig:angular}
\end{figure}

To check whether the excess emission is consistent with a point-like source, 
we show the angular distributions of residual events for 3 energy bands in 
panels {\bf a-c} of Figure \ref{fig:angular}, for the 220 LAT detected sample. 
As a comparison, the angular distributions from the crab (including both the 
pulsar and nebula) for the same energy bands are shown in panels {\bf d-f}. 
We can see that the angular distributions of the excess emission and the 
crab are consistent with each other for $E>0.3$ GeV. For the lowest energy 
bin, slight differences exist which may be due to different spectral shapes 
of them.

\section{Spectrum analysis}

The spectral energy distribution (SED) of GRBs is calculated as
\begin{equation}
\phi(E)=\frac{N_{\text{excess}}}{\Delta E_{\text {bin }} \Delta T A_{\rm eff}},
\label{eq:sed}
\end{equation}
where $N_{\text{excess}}$ represents the number of signal photons within the 
energy bin $\Delta E_{\text{bin}}$ and time bin $\Delta T$. The effective 
area of Fermi-LAT, $A_{\text{eff}}$, depends on the inclination angle 
$(\theta)$, energy $(E)$, and conversion types ({\tt FRONT} or {\tt BACK}) 
of incident photons. For more details one can refer to the
performance\footnote{\url{https://www.slac.stanford.edu/exp/glast/groups/canda/lat_Performance.htm}} 
of the Fermi-LAT detector. The relevant performance parameters are available 
in the calibration 
data\footnote{\url{https://heasarc.gsfc.nasa.gov/FTP/caldb/data/fermi/lat/bcf/ea/aeff_lat.fits}}. 
The systematic uncertainty of the flux calculation due to the effective area 
is estimated to be about $5\%$ \citep{Fermi-LAT:2012fsm}. 

\begin{figure}[!htb]
\centering
\includegraphics[width=0.48\textwidth]{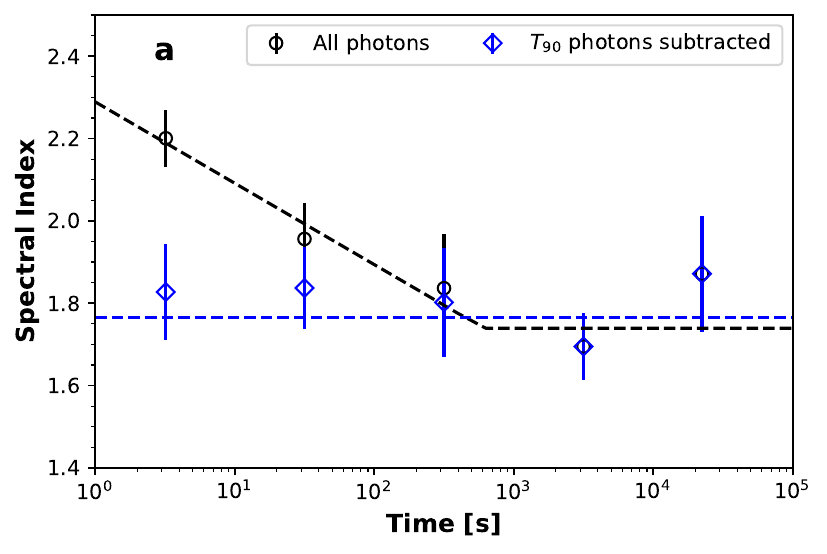}
\includegraphics[width=0.48\textwidth]{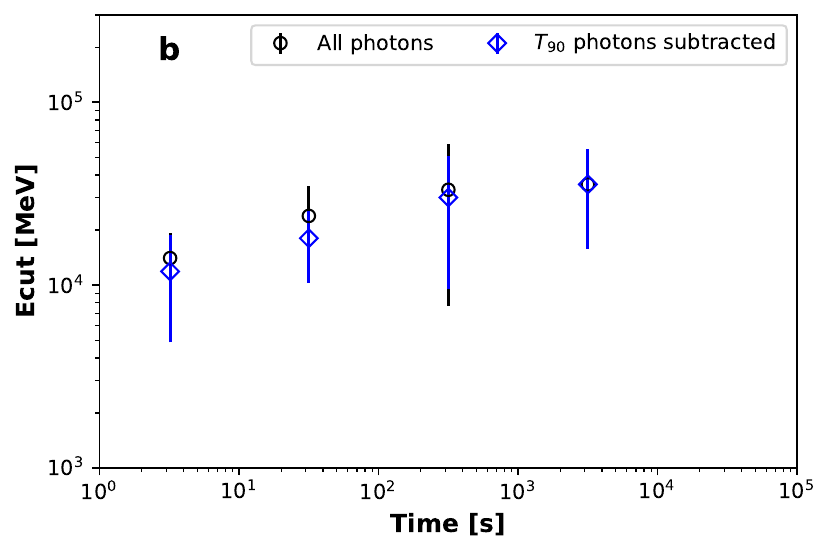}
\caption{The evolution of the spectral index and the cutoff energy. 
Panels {\bf a} and {\bf b} track the temporal evolution of the spectral 
index and the cutoff energy, where black and blue points denote the values 
for the full and $T_{90}$-subtracted samples, respectively.}
\label{fig:index}
\end{figure}

After getting the SED of the stacking GRBs, we use the ECPL model to fit 
the SED. The obtained spectral indices and cutoff energies for different 
data sample (including photons within $T_{90}$ or not) and different time 
bin are shown in Figure \ref{fig:index}. If all the photons 
are taken into account, we find that the spectral index experiences a time 
evolution from soft to hard. After removing photons within $T_{90}$, the 
index keeps almost unchanged with time. The cutoff energy shows a trend 
of increase with time. However, due to the relatively large uncertainties 
of the current data, we cannot draw a firm conclusion about the evolution 
of the spectral cutoff.

\section{Light curves}

To obtain the light curves, we bin the data into different time bins, and 
use a power-law (PL) model of GRBs to fit the data. For the 220 LAT detected 
GRBs, the light curves in two energy bands: 0.1 - 1 GeV and 1 - 100 GeV are 
derived. For the 110 LAT undetected GRBs, only significant emission in the
1 - 10 GeV band is found, and the light curve in this energy band is derived. 

As shown in Figure~\ref{fig:lc}, the light curves shows several segments of 
power laws. A three-piece smoothly broken power law function is employed to 
fit the light curves
\begin{equation}
\phi(t) = \phi_0 t^{\alpha_0} \left[1 + \left(\frac{t}{t_{\rm br,1}}\right)^w\right]^{-\alpha_1/w} 
\left[1 + \left(\frac{t}{t_{\rm br,2}}\right)^w\right]^{-\alpha_2/w},
\label{eq:sbpl}
\end{equation}
where $w$ is the smoothness parameter which is fixed to be 5 in this work. 
For the 110 LAT undetected GRBs, the light curves at early time are not well 
measured due to limited number of photons, and the two-piece smoothly broken 
power law function is enough. Fitting parameters are presented in 
Table \ref{tab:lc}. 

\begin{table}[!ht]
\footnotesize
\centering
\caption{The fitting results of the light curve parameters with $1\sigma$ uncertainties.}
\begin{tabular}{ccccccc}
\hline\hline
& {Energy (GeV)} & {$\alpha_0$} & $t_{\rm br,1}$ (s)& {$\alpha_1$} & {$t_{\rm br,2}$ (s)} & {$\alpha_2$} \\
\hline
220 GRBs & 0.1 - 1  & $0.67\pm0.16$& $1.93\pm3.23$&$0.77\pm0.04$ & $122.20\pm18.10$ & $1.62\pm0.04$ \\
All photons & 1 - 100 &$0.70\pm0.50$ & $1.06\pm0.61$&$0.44\pm0.07$ & $104.58\pm20.17$ & $1.46\pm0.04$  \\
\hline
220 GRBs & 0.1 - 1 &$1.54\pm0.42$ & $1.02\pm0.38$& $0.27\pm0.04$ & $96.91\pm9.4$  & $1.55\pm0.03$   \\
$T_{90}$ subtracted & 1 - 100 & $1.53\pm0.53$& $0.81\pm0.23$&$0.20\pm0.10$  & $93.61\pm16.94$ & $1.46\pm0.04$  \\ 
\hline
110 GRBs & 1 - 10  & -&-&$0.30\pm0.20$ & $8598.97\pm3155.01$ & $3.70\pm1.14$ \\
\multicolumn{1}{c}{All photons}\\
\hline
110 GRBs & 1 - 10 &-& -&$0.16\pm0.22$ & $8134.23\pm3017.92$ & $3.72\pm1.14$  \\
\multicolumn{1}{c}{$T_{90}$ subtracted}\\
\hline
\end{tabular}
\label{tab:lc}
\end{table}

\section{The forward shock model for GRB afterglow}

We use a standard forward shock model to account for the observations of 
this work, speculating that high energy electrons are accelerated by shocks 
due to the interactions of relativistic GRB jets and the circum-burst medium
\citep{Kumar:2014upa}, and the broadband electromagnetic emission is produced 
via the synchrotron and the SSC emission processes 
\citep{Meszaros:1996sv,Sari:1997qe,Wang:2019zbs}. In this model, the emission 
is determined by a set of parameter: the initial Lorentz factor $\Gamma_0$ 
of the jet, the initial isotropic kinetic energy $E_0$, the jet opening angle 
$\theta_j$, the energy partition fractions of the high-energy electrons 
$\epsilon_e$ and magnetic field $\epsilon_B$, the spectral index of electrons 
$p$, the medium density profile as a function of radius $n=n_0(R/R_0)^{-k}$, 
and the redshift of the GRB $z$. 

\begin{figure}[!htb]
\centering
\includegraphics[width=0.48\textwidth]{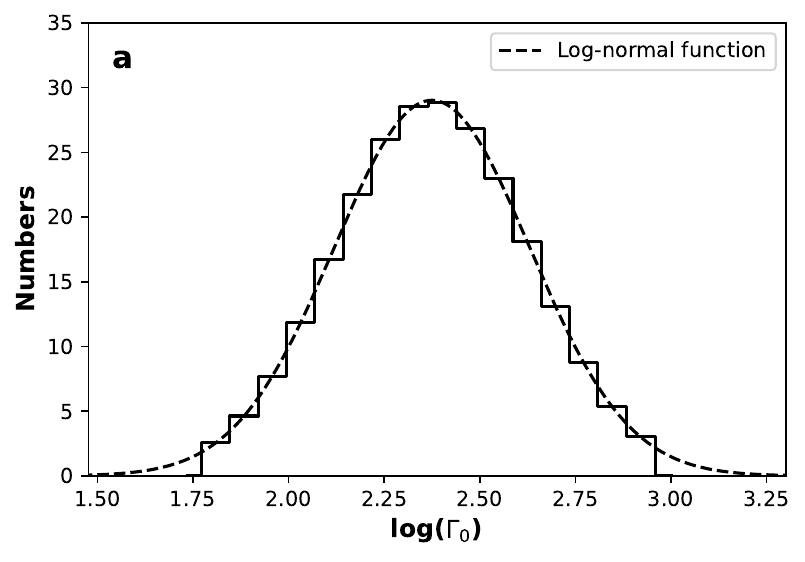}
\includegraphics[width=0.50\textwidth]{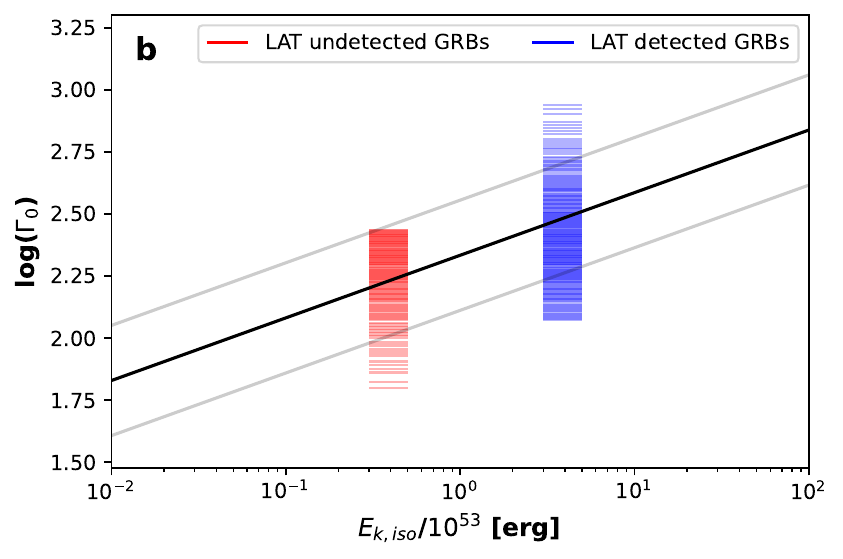}
\caption{Lorentz factor. 
{\bf a} The distribution of $\log(\Gamma_0)$ and the Gaussian fitting curve 
(dashed line). {\bf b} Thick and thin lines show the $\Gamma_0-E_0$ relation 
and the uncertainty ranges given in \citet{Liang:2009zi}. Shaded red and 
blue regions show the parameter regions adopted in the modeling of this work.} 
\label{fig:lognormal}
\end{figure}

A complexity in our case is that we are treating a sample of GRBs instead of 
a single burst. The diversity of the burst parameters need to be taken into 
account. We find that the initial Lorentz factor is a sensitive parameter 
affecting the evolution of the afterglow. We assume a log-normal distribution 
of $\Gamma_0$, with a mean value of $\log(\Gamma_0)=2.37$ and a width of 
$0.26$. The distribution of $\log(\Gamma_0)$ is shown in panel {\bf a} of 
Figure \ref{fig:lognormal}, and is consistent with that given in 
\citet{Liang:2009zi}. The 220 LAT detected sample represents a brighter 
sample in $\gamma$-ray radiation, and a larger Lorentz factor within the 
range of $[130,800]$, a higher initial kinetic energy of $4\times 10^{53}$ 
erg, and a narrower jet opening angle of $0.8^{\circ}$ are assumed. The 
initial Lorentz factor and the initial kinetic energy follows approximately 
the scaling relation of $\Gamma_0\sim E_0^{0.25}$ as found in 
\citet{Liang:2009zi} (see panel {\bf b} of Figure \ref{fig:lognormal}). 
For the 110 LAT undetected sample, the Lorentz factor is restricted to 
the range of $[60,250]$, the initial kinetic energy is set to be 
$4\times 10^{52}$ erg, and the jet opening angle is set to be $4^{\circ}$. 
The other model parameters are fixed as $n_0=1$ cm$^{-3}$, $k=0$, $p = 2.1$, 
$\epsilon_e=0.1$, $\epsilon_B=10^{-4}$, and $z=1$. 

\begin{figure}[!htb]
\centering
\includegraphics[width=0.49\textwidth]{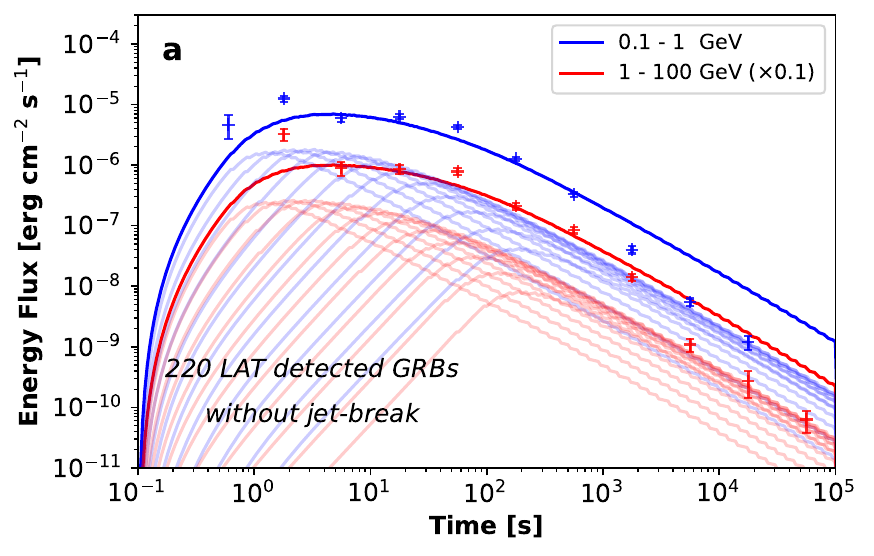}
\includegraphics[width=0.49\textwidth]{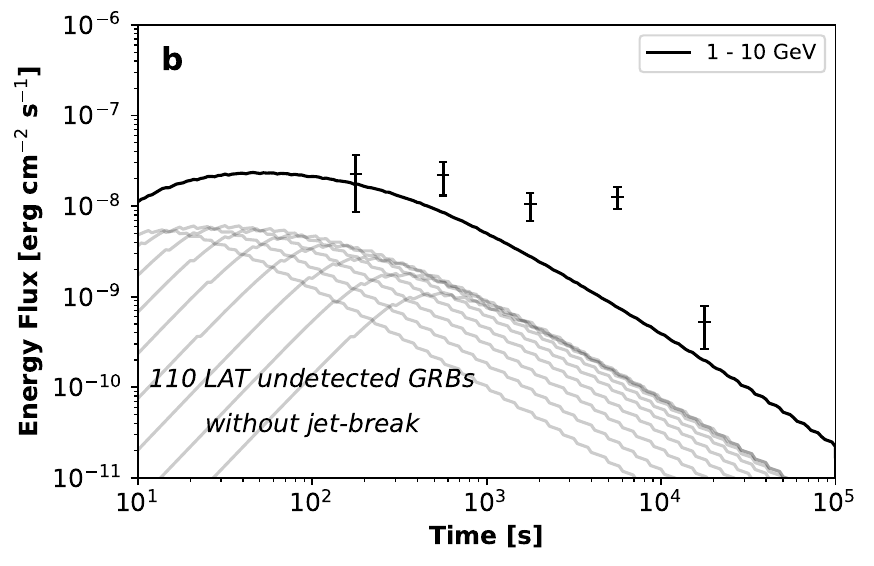}
\caption{Stacking light curves of the GRB populations compared with the 
data, assuming no jet breaks. Panel {\bf a} is for the 220 LAT detected 
sample, and panel {\bf b} is for the 110 LAT undetected sample.}
\label{fig:lc_nojetbreak}
\end{figure}

As can be seen in Figure \ref{fig:model}, the power law index of the second 
segment ($\alpha_1$) of the 220 LAT detected sample is determined by the 
cumulative effect of the GRBs with different Lorentz factors. For individual 
GRB, the initial Lorentz factor can be related with the peak time of the light
curve as \citep{Ghirlanda:2017opl}
\begin{equation}
\Gamma_0\approx 410 (E_0/10^{54}~{\rm erg})^{1/8}(n_0/{\rm cm}^{-3})^{-1/8}
(t_{\rm peak}/10~{\rm s})^{-3/8}.
\end{equation}
For the maximum Lorentz factor of 800, the peak time is about 1.2 s, which 
just corresponds to $t_{\rm br,1}$ of the 220 LAT detected sample. The second 
break at around 100 s corresponds roughly to the minimum Lorentz factors of 
the sample. The jet break effect when the Lorentz factor becomes smaller 
than $1/\theta_j$ also slightly affects the time of the second break 
$t_{\rm br,2}$ and the decay slope after $t_{\rm br,2}$. The jet break time 
can be estimated as \citep{Wu:2004jx}
\begin{equation}
t_{\rm jet}\approx 1210~(\theta_j/1^{\circ})^{8/3}(E_0/10^{54}~{\rm erg})^{1/3}(n_0/{\rm cm}^{-3})^{-1/3}
~{\rm s}.
\end{equation}
Figure \ref{fig:lc_nojetbreak} show the light curves assuming that there 
is no jet break effect, which are inconsistent with the data at late time. 
Therefore, our data suggest the existence of jet breaks in general.

\section{X-ray light curves from Swift/XRT}

The light curve of the 110 LAT undetected GRBs cannot be well explained by 
the above model. We expect that there might be late-time energy injection 
effect in the data. In this case, the light curve of the X-ray afterglow 
would show a shallower-than-normal decay behavior \citep{Zhang:2005fa}. 
We therefore collect the Swift/XRT light curves of the two samples to 
investigate whether there are differences of their X-ray light curves. 

\begin{table}[!ht]
\footnotesize
\centering
\caption{The classification of Swift/XRT light curves for the samples.}
\begin{tabular}{ccccccc}
\hline\hline
& {No break} & {One break} & {Canonical} & {Oddball} & {TBD} & {None} \\
\hline
220 LAT detected GRBs & 57  & 11 & 15 & 15 & 10 & 112 \\
\hline
110 LAT undetected GRBs & 43  & 23 & 24 & 14 & 6 & - \\
\hline
\end{tabular}
\label{tab:XRT}
\end{table}

\begin{figure}[!htb]
\centering
\includegraphics[width=0.49\textwidth]{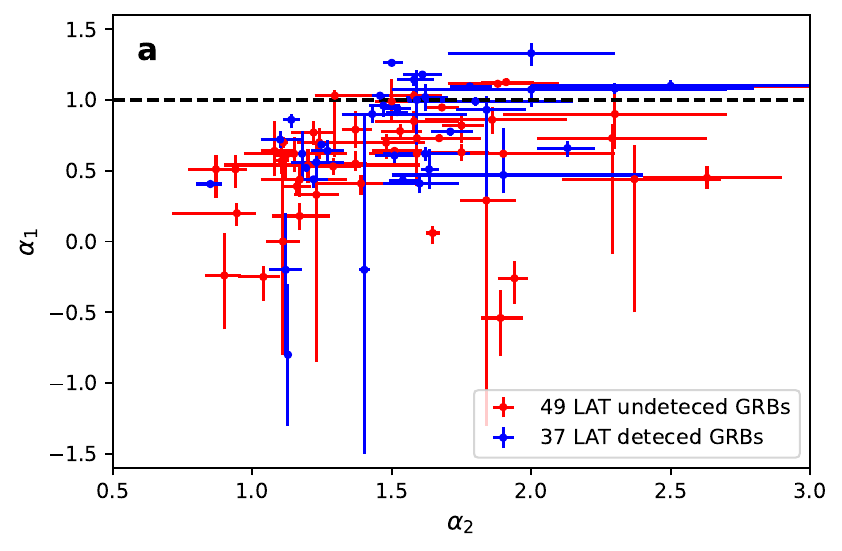}
\includegraphics[width=0.49\textwidth]{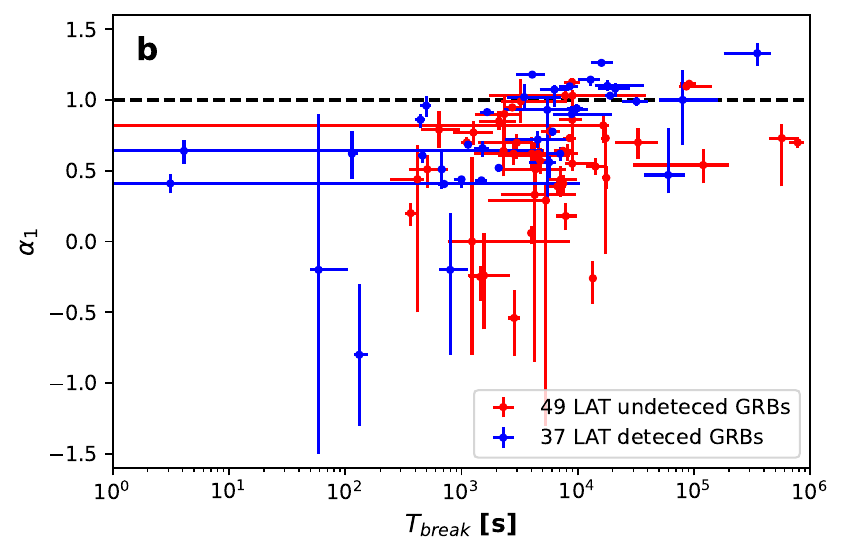}
\caption{Statistical analysis of Swift/XRT light curves. Panel {\bf a} 
shows the distributions of the two decay indices derived from the Swift/XRT 
light curves of the sample. Panel {\bf b} presents the distribution of the 
first decay index against the subsequent break time in the Swift/XRT light 
curves.}
\label{fig:alpha1}
\end{figure}

We classify the Swift/XRT light curves into five categories as defined in 
\citet{Evans:2008wp}: ``No break'', ``One break'', ``Canonical'', ``Oddball'', 
and cases with insufficient data (``TBD'' or ``None''). The results are 
summarized in Table \ref{tab:XRT}. The categories that may be used to 
identify possible energy injection features include the 
``Canonical'', ``One break'', and ``Oddball'' cases. It shows that 37 out 
of the 108 GRBs with Swift/XRT data of the 220 LAT-detected GRBs belong to 
these three categories, and for the 110 LAT undetected sample, 49 out of 
110 belong to these three categories. The fraction is slightly higher for 
the 110 LAT undetected sample. We further show the slope parameters and 
the break time of the Swift/XRT light curves in Figure \ref{fig:alpha1}. 
In panel {\bf a} the power-law indices $\alpha_1$ and $\alpha_2$ are shown, 
and in panel {\bf b} the slope $\alpha_1$ versus 
the break time are shown. Here $\alpha_1$ and $\alpha_2$ are the temporal 
decay indices before and after the break, respectively, and $T_{\rm break}$ 
is the break time (when there are multiple breaks, the parameters describe 
segments II and III in \citet{Zhang:2005fa} are adopted). 
It can be seen that the majority of the 49 LAT undetected GRBs have 
$\alpha_1 < 1$, indicating that there might be energy injection processes 
for these GRBs. As a comparison, the 37 LAT detected GRBs have $\alpha_1$ 
values either bigger or smaller than 1. The break time of the 49 LAT 
undetected GRBs are mostly within $10^3$ and $10^4$ s, while that of the 
37 LAT detected GRBs are more broadly distributed. These properties provide 
support for the existence of energy injection features for the 110 LAT 
undetected sample. 

\bibliography{refs}
\bibliographystyle{apj}

\end{document}